# A hybrid proper orthogonal decomposition and diffusion framework for reduced-order forecasting of turbulent flow dynamics

**Authors:** Rodrigo Abadia-Heredia[1], Xiangrui Zou[1*], Manuel Lopez-Martin[2], Petros Koumoutsakos[3], Soledad Le Clainche[1*]

[1] ETSI Aeronautica y del Espacio, Universidad Politécnica de Madrid, Plaza Cardenal Cisneros, 3, Madrid 28040, Spain.

[2] Advanced Mathematics Group, Repsol Technology Lab, Repsol S.A., Madrid 28935, Spain.

[3] Computational Science and Engineering Laboratory, Harvard University, 29 Oxford Street, Cambridge, MA 02138, USA.

**Corresponding Authors**:

Xiangrui Zou, Email: x.zou@upm.es

Tel: +34 600 67 89 35, Fax: +34 600 67 89 35

ETSI Aeronautica y del Espacio, Universidad Politécnica de Madrid, Plaza Cardenal Cisneros, 3, Madrid 28040, Spain.

Soledad Le Clainche, Email: soledad.leclainche@upm.es

Tel: +34 910 67 58 53, Fax: +34 910 67 58 53

ETSI Aeronautica y del Espacio, Universidad Politécnica de Madrid, Plaza Cardenal Cisneros, 3, Madrid 28040, Spain.

**Abstract**

Forecasting turbulent flow dynamics requires a balance between predictive fidelity and computational efficiency. Diffusion-based generative models can represent complex spatiotemporal dynamics, but their application to high-dimensional turbulent flows remains computationally expensive. In contrast, proper orthogonal decomposition (POD) provides compact, physically interpretable reduced-order representations, although aggressive modal truncation can remove relevant flow structures. This work introduces a hybrid reduced-order generative forecasting framework that combines POD with Generative Learning of Effective Dynamics (G-LED). The method performs temporal prediction in a physics-based modal space and uses diffusion-based reconstruction to recover physically meaningful flow-field representations. It is assessed using experimental measurements of the turbulent wake behind a circular cylinder. Three configurations are compared: full-field G-LED, global POD-G-LED, and localized POD-G-LED. Full-field G-LED provides the highest fidelity, preserving richer vorticity fluctuations and more consistent turbulent kinetic energy distributions, but requires approximately 17 h for diffusion-model training, 7 h for Transformer training, and 3 min to predict 100 future snapshots. By transferring prediction to a reduced POD space, global POD-G-LED reduces these costs to approximately 8 h, 2 h, and 50 s, respectively, while retaining dominant wake organization and coherent energetic structures. A localized POD-G-LED formulation assigns different modal resolutions to distinct wake regions and improves vorticity statistics and energetic distributions relative to the global reduced-order configuration. These results show that coupling physics-based modal representations with diffusion-based generative reconstruction offers an effective route to efficient turbulent-flow forecasting and provides a bridge between low-dimensional flow representations and physically meaningful reconstruction.

## 1. Introduction

Flow-field prediction and surrogate modeling are essential for the study of many fluid-mechanics problems [1]. With the rapid development of computational techniques, both computational fluid dynamics (CFD) and deep learning (DL) models have been increasingly applied in fluid mechanics [2–5]. However, accurate long-term prediction remains very challenging for high-dimensional turbulent flows, because their dynamics involve a wide range of strongly coupled spatial and temporal scales, and the small-scale structures are highly irregular and stochastic [6,7]. These small-scale structures therefore pose a major challenge for turbulence prediction.

In recent years, diffusion-based generative models have attracted growing attention in spatiotemporal data modeling and the prediction of complex dynamical systems [8,9]. Previous studies have shown that diffusion models have been widely used for forecasting, generation, imputation, and anomaly detection in time-series and spatiotemporal data, mainly because they can explicitly characterize complex data distributions and support conditional and probabilistic prediction. In the prediction of complex dynamical systems, DYffusion [8,10] is one of the representative works that explicitly introduced diffusion ideas into spatiotemporal forecasting. Instead of only performing static generation, it coupled the diffusion process with system dynamics for probabilistic multi-step prediction. In fluid mechanics, Lienen et al. [11] showed at ICLR 2024 that diffusion models can be directly used for three-dimensional flow-field generation, demonstrating the feasibility of generative modeling for 3D turbulence simulation. Du et al. [12] proposed CoNFiLD, which combines conditional neural fields with latent diffusion for high-fidelity and stochastic spatiotemporal turbulence generation in complex three-dimensional domains, and highlighted the advantage of latent diffusion in reducing memory usage and generation cost. In addition, Gao et al. [13] proposed the generative learning of effective dynamics (G-LED) framework, which further combines low-dimensional effective-dynamics prediction with high-dimensional diffusion-based reconstruction. In this framework, the macro dynamics are first predicted on a low-dimensional manifold, and the high-dimensional states are then recovered by a Bayesian diffusion decoder. This provides a new way to balance fidelity and efficiency in the statistical prediction of high-dimensional complex systems, especially turbulent flows.

Although G-LED has shown some advantages in predicting small-scale structures, its training and inference often require substantial computational resources, which limits its efficient use in practice, especially for online deployment and rapid iteration. Reduced-order modeling and data-driven surrogate models have become increasingly important in CFD [14,15]. Reduced-order models replace the repeated evolution of high-dimensional flow fields with a low-dimensional representation, thereby significantly reducing the computational cost while retaining the main physical features as much as possible. Among many reduced-order techniques, proper orthogonal decomposition (POD) [16] plays a central role because it extracts an orthonormal basis ordered by energy and provides a compact representation of coherent flow structures. In many wake and shear-flow problems, only a limited number of POD modes are needed to capture the dominant large-scale organization of the flow, whereas higher-order modes correspond to increasingly fine, irregular, and low-energy flow components. Because of this energy-ordering property, POD is especially attractive for machine-learning-assisted flow prediction. Instead of learning directly on high-dimensional spatiotemporal tensors, the model only needs to learn the evolution of a much smaller number of POD coefficients. The resulting latent representation is not only compact, but also physically interpretable.

POD has been widely used in flow prediction [17–22]. In early studies, POD was mainly combined with Galerkin projection to construct classical reduced-order models. In such methods, the Navier–Stokes equations are projected onto a modal subspace, and the flow evolution is solved directly in the reduced space, so that the main dynamical features can be obtained at a much lower cost than full CFD. To overcome the dependence of conventional POD-Galerkin models on the governing-equation structure and closure modeling, many recent studies have shifted toward non-intrusive predictive frameworks that combine POD with machine learning or deep learning. For example, the predictive reduced-order model proposed by Abadía-Heredia et al. [20], which combines POD with deep learning, has been shown to be effective for representative fluid-flow problems. Similarly, the coupling of POD with sequence models such as long short-term memory (LSTM) has also been used for deterministic and statistical prediction of unsteady wakes and complex turbulent flows [23,24]. In these approaches, POD is used to extract low-dimensional modes and their temporal coefficients from flow snapshots, while neural networks, LSTMs, or other temporal propagators are used to learn the evolution of the POD coefficients. Such hybrid frameworks can preserve the physical interpretability of POD while improving long-term prediction capability and generalization across different operating conditions.

However, the limitations of POD in flow prediction are also clear. Because POD modes are ranked by energy, the higher-order low-energy modes often correspond to smaller, more irregular, and more difficult-to-predict flow structures. Once these modes are truncated, the corresponding information is irreversibly lost. As a result, POD naturally involves a trade-off between efficiency and fidelity: retaining fewer modes can greatly reduce the computational cost, but may weaken the ability to represent small-scale structures and complex statistical features; retaining more modes can improve reconstruction accuracy, but also increases the difficulty of model training and prediction. When finer structures are to be captured, more POD modes can be retained, and this is precisely where the G-LED model may become useful in the subsequent prediction stage.

Motivated by these considerations, this work investigates the integration of physics-based reduced-order modeling and diffusion-based generative forecasting. The central hypothesis is that POD can provide a compact and physically interpretable representation of turbulent flow dynamics, while G-LED can enhance the predictive capability of this reduced-order space through diffusion-based generative reconstruction. In this way, the forecasting problem is shifted from high-dimensional flow fields to low-dimensional modal dynamics without completely sacrificing the spatial richness of the reconstructed flow.

The contributions of this work are fourfold. First, a hybrid POD-G-LED framework is proposed for reduced-order generative forecasting of turbulent cylinder wakes. Second, the study quantifies the trade-off between predictive fidelity and computational efficiency when diffusion-based forecasting is performed in a reduced-order modal space. Third, a localized POD-G-LED formulation is introduced, allowing different wake regions to be represented with different modal resolutions according to their local dynamical complexity. Fourth, the results demonstrate that diffusion models can be effectively combined with physics-based modal decompositions to reconstruct physically meaningful turbulent-flow representations at a fraction of the computational cost required by direct full-field generative forecasting.

The remainder of this paper is organized as follows. Section 2 introduces the methodology used in this study, including the G-LED model, the POD-G-LED coupling strategy, and the evaluation metrics. Section 3 presents the prediction results obtained with different strategies and systematically compares their performance in terms of accuracy and efficiency. Section 4 discusses the results, with particular attention to POD truncation, localized partition strategies, and their influence on flow-field fidelity and computational cost. Finally, Section 5 summarizes the main conclusions and outlines possible directions for future work.

## 2. Methods

### 2.1. Flow data and prediction task

To assess the predictive performance of the proposed model on more complex dynamics, the experimental dataset of turbulent flow past a circular cylinder is employed [25]. This dataset consists of two-dimensional velocity fields obtained from time-resolved particle image velocimetry (TR-PIV) measurements of the three-dimensional wake behind a circular cylinder. The experiments were conducted in the L10 low-speed wind tunnel at the von Kármán Institute; further details can be found in Mendez et al. [25].

The full dataset contains 13,200 snapshots and spans approximately 4.5 time units, during which the free-stream velocity transitions between two steady operating conditions. The first 4,000 snapshots correspond to the initial steady regime at Re = 4000, the next 4,000 snapshots describe the transient regime, and the final 5,200 snapshots correspond to the second steady regime at Re = 2600, as illustrated in Fig. 1. In the present work, the first 3,200 snapshots from the second steady regime are used for training, and the subsequent 100 snapshots are used for prediction and evaluation.

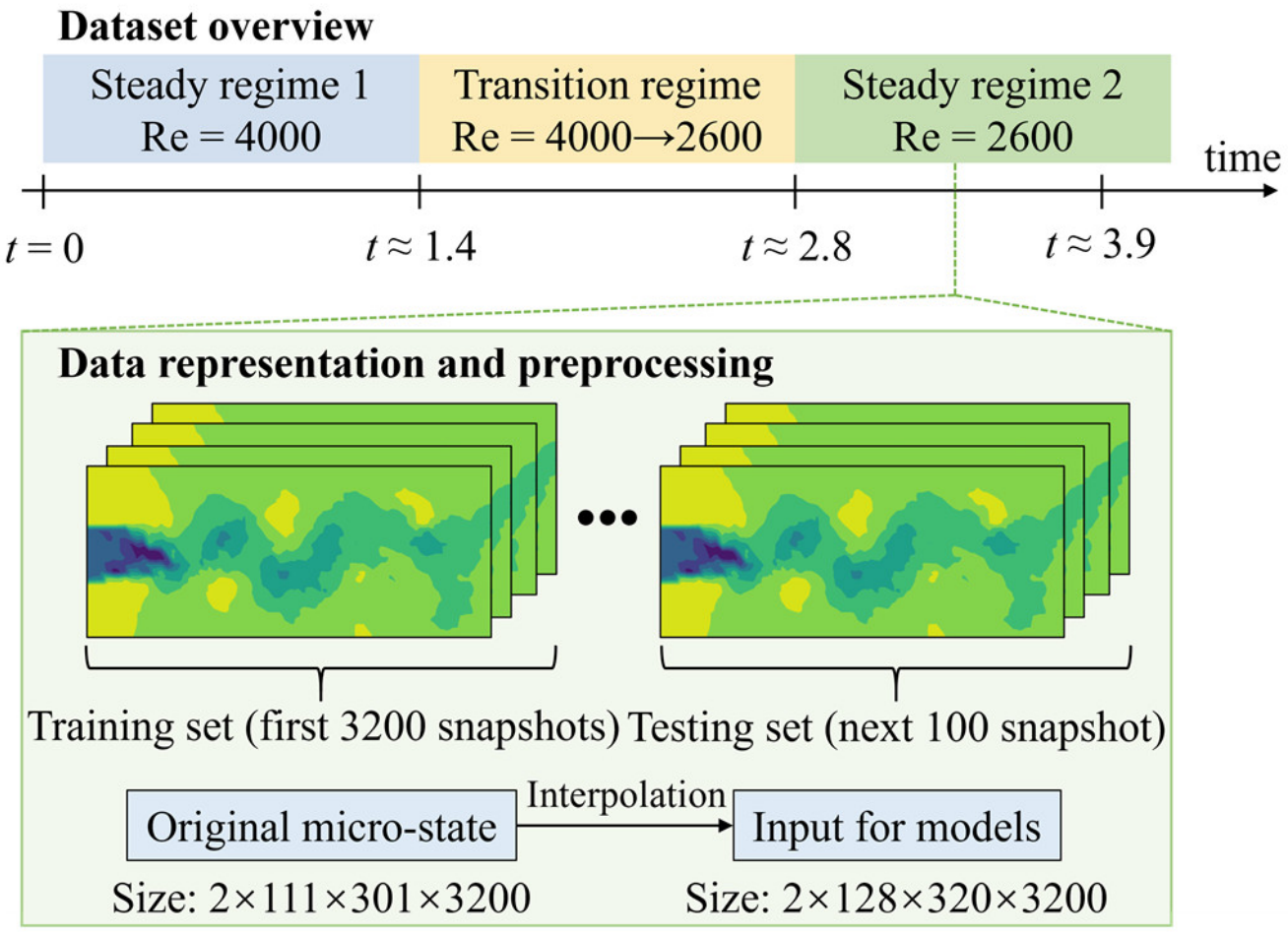

**Fig. 1.** Schematic illustration of the dataset organization and preprocessing procedure. The full dataset is divided into two steady regimes and one transient regime. In this study, the first 3,200 snapshots from the second steady regime are used for training, and the next 100 snapshots are used for prediction and evaluation. The original velocity fields are interpolated from 2×111×301 to 2×128×320 to satisfy the architectural requirements of the U-Net model.

The original micro-state is the instantaneous velocity field. In the direct full-field configuration, each snapshot is represented as

$$\mathbf{v}_t \in \mathbb{R}^{C \times N_x \times N_y}, \tag{1}$$

where $C = 2$ represents the two measured velocity components, and $N_x$ and $N_y$ are the numbers of grid points in the streamwise and transverse directions, respectively. For the current dataset, $N_x = 111$ and $N_y = 301$ [25]. The total number of spatial points is therefore $N = N_x \times N_y$. To satisfy the architectural requirements of the U-Net [26] used in the diffusion model, the velocity fields are linearly interpolated to

$$\mathbf{v}_t \in \mathbb{R}^{2 \times 128 \times 320}. \tag{2}$$

This interpolation is a practical preprocessing step required by the repeated downsampling and upsampling operations in the convolutional network architecture.

### 2.2. Full-field G-LED method

A schematic illustration of the G-LED method is provided in Fig. 2. G-LED separates prediction into two stages: a low-dimensional temporal forecast of a coarse representation and a conditional generative reconstruction of the corresponding high-dimensional state. In this work, the terminology of micro-state and macro-state follows the G-LED formulation. The micro-state denotes the detailed flow representation to be reconstructed by the generative model. In the full-field configuration, it corresponds to the interpolated velocity snapshot $\mathbf{v}_t \in \mathbb{R}^{2\times128\times320}$. The macro-state denotes a lower-dimensional coarse representation obtained from the micro-state through a fixed non-trainable encoder. In the full-field configuration, this macro-state is $\mathbf{z}_t \in \mathbb{R}^{2\times32\times32}$. The Transformer predicts the temporal evolution of the macro-state, while the diffusion decoder reconstructs the corresponding micro-state.

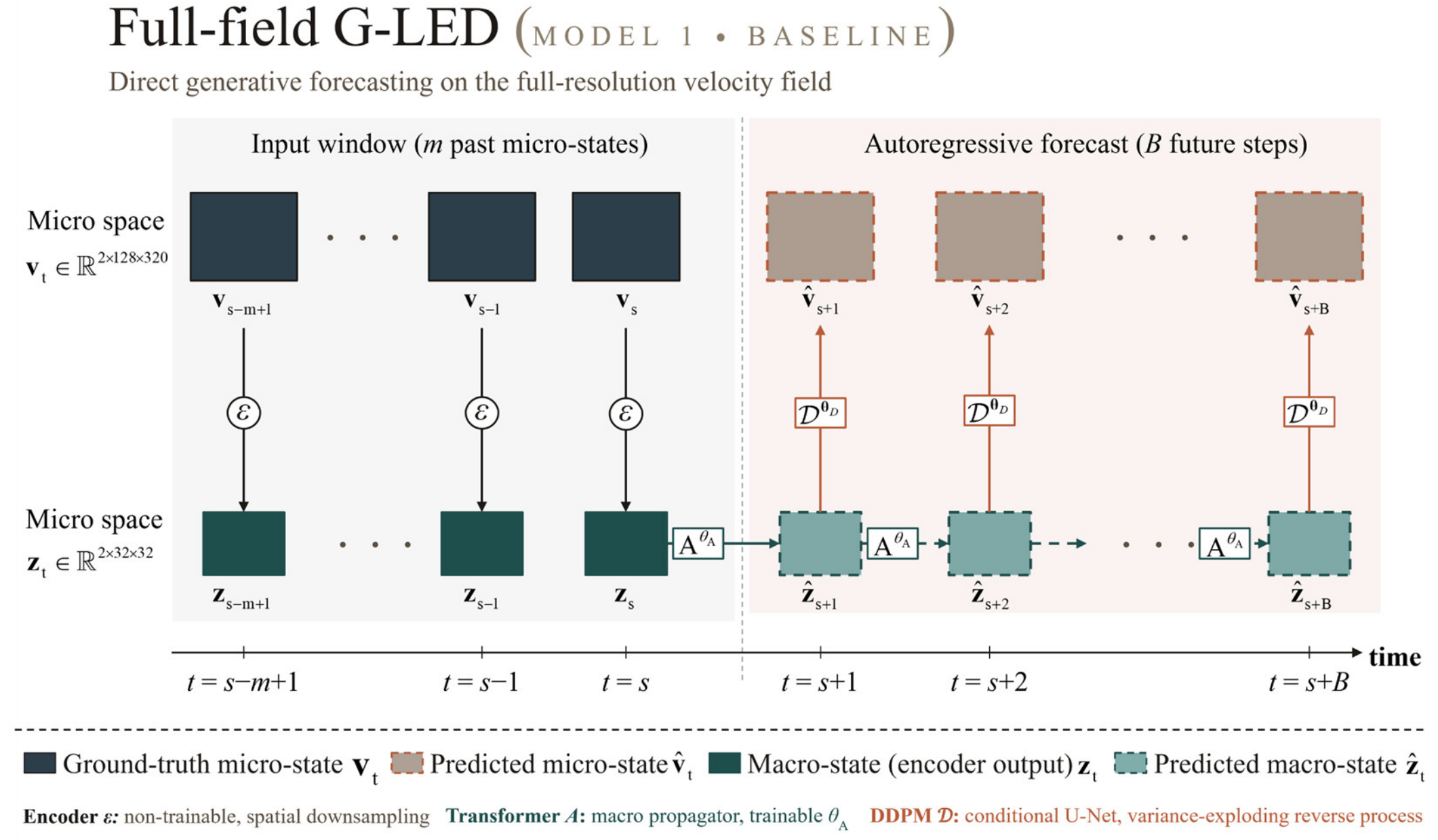

**Fig. 2.** Schematic illustration of the full-field G-LED configuration. The input consists of a temporal window of $m$ high-dimensional micro-states $\mathbf{v}_t \in \mathbb{R}^{2\times128\times320}$, corresponding to full-field velocity snapshots. A fixed non-trainable encoder E, implemented as spatial downsampling, maps each micro-state to a lower-dimensional macro-state $\mathbf{z}_t \in \mathbb{R}^{2\times32\times32}$. The Transformer propagator advances the macro-state autoregressively over B future time steps. The conditional diffusion decoder then reconstructs the corresponding predicted high-dimensional micro-states $\hat{\mathbf{v}}_t$ from the predicted macro-states.

Let the micro-state at time $t$ be denoted by $\mathbf{v}_t$, and the physical evolution over one time step $\Delta t$ is

written as

$$\mathbf{v}_{t+\delta t} = \mathbb{F}\left(\mathbf{v}_t\right), \tag{3}$$

where $\mathbb{F}: \mathbb{R}^{C \times N} \to \mathbb{R}^{C \times N}$ denotes the numerical operator that advances the solution by a predefined time step $\delta_t$. This iterative process requires an initial condition, $\mathbf{v}_0$. After propagating the solution over $S$ time steps, the resulting dataset, **D**, is given by

$$\mathbf{D} = \left[\mathbf{v}_0, \mathbf{v}_1, \ldots, \mathbf{v}_{S\text{-}1}\right], \mathbf{D} \in \mathbb{R}^{C \times N \times S}. \tag{4}$$

In practical scenarios, $N \gg 1$, leading to a substantial number of degrees of freedom, often on the order of thousands to millions, which significantly increases the computational cost. Instead of learning this mapping directly in the original high-dimensional space, G-LED [13] introduces a lower-dimensional macro-state $\mathbf{z}_t \in \mathbb{R}^{C \times N_z}$, obtained through a non-trainable encoder, given as

$$\mathbf{z}_t = \varepsilon\left(\mathbf{v}_t\right), \tag{5}$$

where $\mathbf{v}_t$ and $\mathbf{z}_t$ are referred to as the micro and macro representations of the solution at time $t$, respectively. In the present full-field configuration, $\varepsilon$ is implemented as a fixed non-trainable spatial downsampling operator that maps the interpolated velocity field from 2×128×320 to 2×32×32. This encoder contains no trainable parameters and is used only to define the coarse macro-state evolved by the Transformer. For the present study, $\mathbf{z}_t$ is given as

$$\mathbf{z}_t \in \mathbb{R}^{2 \times 32 \times 32}. \tag{6}$$

The macro representation $\mathbf{z}_t$ is characterized by the absence of small-scale structures, which are typically associated with near-chaotic dynamics. To forecast the temporal evolution of the system, a macro propagator $A^{\theta_A}$, parameterized by trainable weights $\theta_A$ and based on the Transformer architecture, is employed

$$A^{\theta_A}: \begin{cases} \left(z_0, z_{\Delta t}, \ldots, z_{(n_t-1)\Delta t}\right) \to z_{n_t \Delta t}, if\ n_t \le N_t \\ \left(z_{(n_t-N_t)\Delta t}, z_{(n_t-N_t+1)\Delta t}, \ldots, z_{(n_t-1)\Delta t}\right) \to z_{n_t \Delta t}, if\ n_t > N_t \end{cases}, \tag{7}$$

where $N_t$ is the temporal lookback window and $\theta_A$ denotes the trainable parameters. According to Refs. [27,28], the optimal trainable weights $\theta_A^*$ of the propagator are obtained by minimizing the MSE, given

as

$$\theta_{\mathrm{A}}^{*}=\arg\min_{\theta_{\mathrm{A}}}\sum_{n_{\mathrm{t}}\in\mathbf{D}_{\mathrm{train}}}\sum_{i=1}^{N_{\mathrm{t}}}\left\|z_{\mathrm{n_t}\Delta\mathrm{t}}-\mathrm{A}^{\theta_{\mathrm{A}}}\left(z_{(\mathrm{n_t}-\mathrm{i})\Delta\mathrm{t}},\ldots,z_{(\mathrm{n_t}-1)\Delta\mathrm{t}}\right)\right\|_{2}^{2}, \tag{8}$$

where $\mathbf{D}_{\mathrm{train}}$ denotes the subset of temporal snapshots used for training the propagator. The full dataset $\mathbf{D}$ is partitioned into disjoint training and testing sets, such that $\mathbf{D} = \mathbf{D}_{\mathrm{train}} \cup \mathbf{D}_{\mathrm{test}}$, in which $\mathbf{D}_{\mathrm{test}}$ is reserved exclusively for model evaluation.

The second stage reconstructs the corresponding micro-dynamics conditioned on the macro trajectory. This is done with a denoising diffusion probabilistic model (DDPM) implemented through a conditional U-Net [26]. This generative framework maps a temporal window of macro inputs, $z_{\mathrm{t:t+B\Delta t}}$, to the corresponding micro outputs, $\mathbf{v}_{\mathrm{t:t+B\Delta t}}$. A DDPM is a generative framework composed of two sequential stages: a forward (noising) process and a reverse (denoising) process. The forward process progressively corrupts the micro data by adding Gaussian noise over $N_{\mathrm{e}}$ steps according to a Markov chain

$$\mathbf{v}_{\mathrm{t:t+B\Delta t}}\equiv e^{0}\rightarrow e^{1}\rightarrow\ldots\rightarrow e^{N_{\mathrm{e}}}, \tag{9}$$

$$e^{\mathrm{i}}\sim\mathcal{N}\left(e^{\mathrm{i}};\sqrt{1-\sigma_{\mathrm{i}}^{2}}e^{\mathrm{i}-1},\sigma_{\mathrm{i}}^{2}\mathbf{I}\right),i=1,\ldots,N_{\mathrm{e}}, \tag{10}$$

where $\{\sigma_2^{\mathrm{i}}\}_{\mathrm{i}=1}^{\mathrm{N}\epsilon}$ is a pre-defined variance schedule. This schedule is strictly monotonically increasing [29] and it is chosen such that early noise levels are minimal, preserving the key features of the original data (i.e., $e^0 \simeq e^1$), while later stages result in a fully randomized sample (i.e., $e^{\mathrm{N}\epsilon} \sim \mathcal{N}(0, \mathbf{I})$) with $\sigma_{\mathrm{N}\epsilon} \gg \|\mathbf{v}_{\mathrm{t:t+B\Delta t}}\|_2$ [30]. This schedule of variances is computed by

$$\sigma_{i}=\left(\sigma_{N_{\epsilon}}^{\rho^{-1}}+\frac{N_{\epsilon}-i}{N_{\epsilon}-1}\left(\sigma_{1}^{\rho^{-1}}-\sigma_{N_{\epsilon}}^{\rho^{-1}}\right)\right)^{\rho},\quad\text{for } i=1,2,\ldots,N_{\epsilon}, \tag{11}$$

where $N_\epsilon$ is a hyperparameter of the diffusion model selected based on previous works [29,31,32]. Here, $\rho = 7$ is used to obtain a controlled nonlinear progression of the variance schedule [33].

On the other hand, the reverse process starts from a white noise sample $\epsilon \sim \mathcal{N}(0, \mathbf{I})$ and it learns to gradually remove the noise until recovering the original micro dynamics $\mathbf{v}_{\mathrm{t:t+B\Delta t}}$. The reverse process used for G-LED is known as the variance-exploding diffusion process [34], where the reverse process

is tractable [30,35]

$$p(\mathbf{e}^i | \mathbf{v}_{t:t+B\Delta t}, \mathbf{e}^{i+1}) = \mathcal{N}\left( \frac{\sigma_{i+1}^2 - \sigma_i^2}{\sigma_{i+1}^2} \mathbf{v}_{t:t+B\Delta t} + \frac{\sigma_i^2}{\sigma_{i+1}^2} \mathbf{e}^{i+1}, \frac{(\sigma_{i+1}^2 - \sigma_i^2)\sigma_i^2}{\sigma_{i+1}^2} \mathbf{I} \right). \tag{12}$$

However, the micro dynamics $\mathbf{v}_{t:t+B\Delta t}$ are unknown, and therefore they are approximated by a deep neural network (DNN), $\hat{\mathbf{S}}_{\theta_D} : \mathbb{R}^{C \times N_z \times B} \times \mathbb{R}^{C \times N \times B} \times \mathbb{N} \to \mathbb{R}^{C \times N \times B}$, parameterized by trainable weights $\boldsymbol{\theta}_D$. This DNN receives the macro dynamics as input and is trained to approximate the corresponding micro dynamics. The reverse process is therefore expressed as

$$p(\mathbf{e}^i | \mathbf{v}_{t:t+B\Delta t}, \mathbf{e}^{i+1}) \approx p(\mathbf{e}^i | \mathbf{z}_{t:t+B\Delta t}, \mathbf{e}^{i+1}, i) = \mathcal{N}\left( \frac{\sigma_{i+1}^2 - \sigma_i^2}{\sigma_{i+1}^2} \mathcal{D}^{\boldsymbol{\theta}_D}(\mathbf{z}_{t:t+B\Delta t}, \mathbf{e}^{i+1}, i) + \frac{\sigma_i^2}{\sigma_{i+1}^2} \mathbf{e}^{i+1}, \frac{(\sigma_{i+1}^2 - \sigma_i^2)\sigma_i^2}{\sigma_{i+1}^2} \mathbf{I} \right). \tag{13}$$

The DNN employed within G-LED is a U-Net architecture [26]. The optimal parameters of the diffusion model, $\boldsymbol{\theta}_D^*$, are learned by minimizing the Kullback–Leibler (KL) divergence between the true and approximate reverse diffusion processes, denoted in Eqs. (12) and (13). Under standard assumptions, as detailed by Refs. [35,36], this minimization yields the loss function

$$\boldsymbol{\theta}_D^* = \arg\min_{\boldsymbol{\theta}_D} \sum_{n_t \in \mathbf{D}_{train}} \mathbb{E}_{i,\mathbf{e}^i} \left\| \mathcal{D}^{\boldsymbol{\theta}_D}(\mathbf{z}_{t:t+B\Delta t}, \mathbf{v}_{t:t+B\Delta t}^i, i) - \mathbf{v}_{t:t+B\Delta t} \right\|_2, \tag{14}$$

where $\hat{\mathbf{S}}_{\theta_D}$ denotes the network's estimate of the noise at diffusion step $i$, conditioned on the macro inputs. The G-LED model results from coupling the macro propagator with the diffusion decoder, as originally proposed by Gao et al. [13], for forecasting high-dimensional spatiotemporal dynamics.

### 2.3. Global POD-G-LED method

A key methodological extension considered in the work is the introduction of POD before the generative model. This changes both the representation of the flow and the interpretation of the forecasting task. Instead of applying G-LED directly to high-dimensional velocity snapshots, the model is trained and evaluated in a POD coefficient space. Fig. 3 shows a schematic illustration of the global POD-G-LED framework.

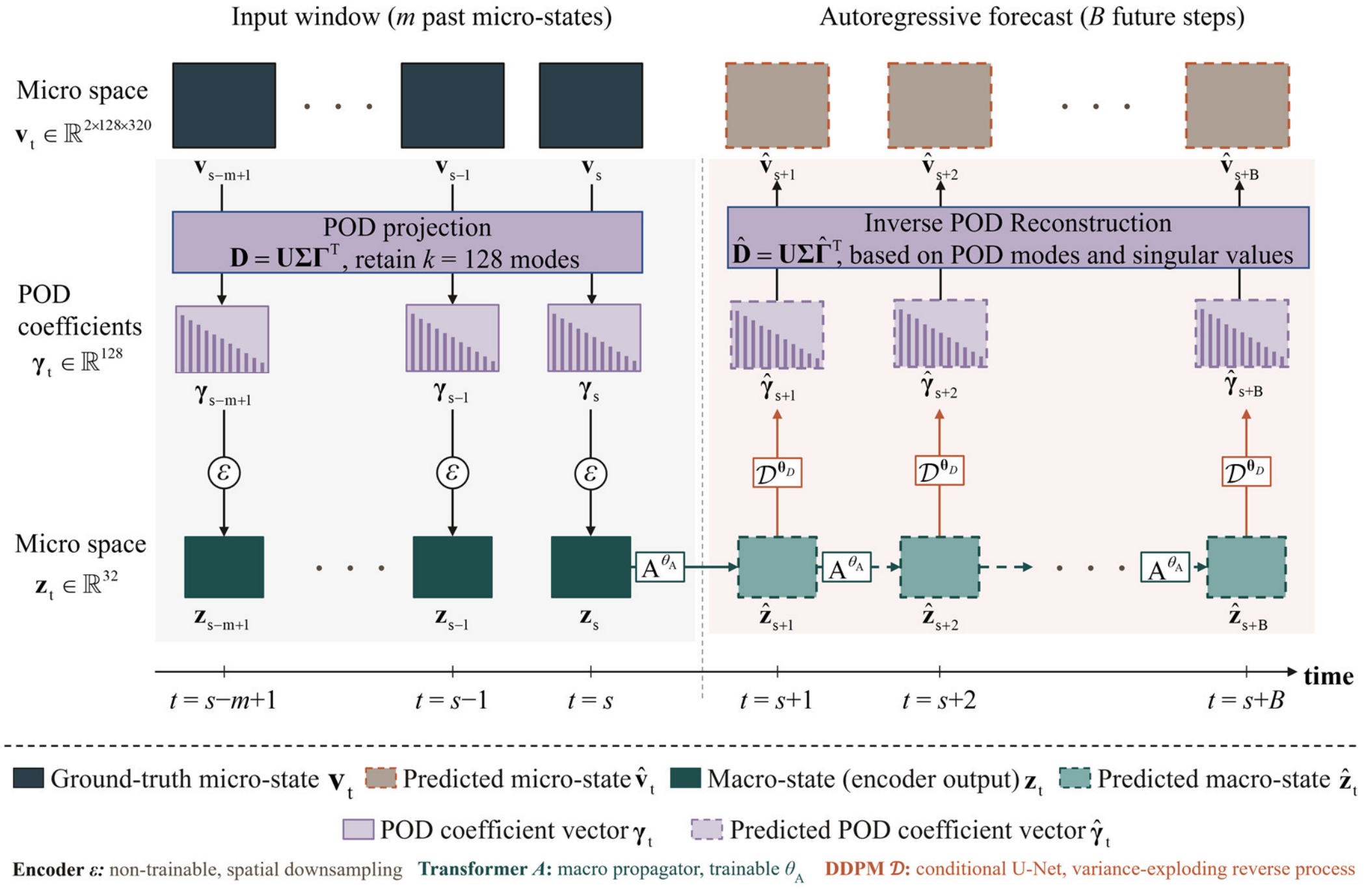

**Fig. 3.** Schematic illustration of the global POD-G-LED configuration. The original velocity snapshots vt are first projected onto a global POD basis, yielding retained POD coefficient vectors $\mathbf{a}_t \in \mathbb{R}^{128}$. These coefficient vectors define the reduced-order micro-states used in the POD-G-LED configuration. A fixed encoder further maps $a_t$ to a lower-dimensional macro-state $\mathbf{z}_t \in \mathbb{R}^{32}$, which is propagated in time by the Transformer. The conditional diffusion decoder reconstructs the predicted POD coefficient vector $\hat{\mathbf{a}}_t$ from the predicted macro-state, and the final velocity field $\hat{\mathbf{v}}_t$ is obtained by inverse POD reconstruction using the retained modes.

In the global POD-G-LED configuration, POD is applied to the training snapshots to obtain a reduced-order representation of the flow. The snapshot matrix is decomposed as

$$\mathbf{D} = \mathbf{U}\boldsymbol{\Sigma}\boldsymbol{\Gamma}^{\mathrm{T}}, \tag{15}$$

where $\mathbf{D}$ denotes the snapshot matrix, $\mathbf{U}$ is the matrix that contains the spatial POD modes, $\boldsymbol{\Sigma}$ is a diagonal matrix containing the singular values, and $\boldsymbol{\Gamma}$ contains the temporal POD coefficients. In the present study, the 128 most energetic modes are retained. The number of retained modes is selected as a compromise between reconstruction fidelity and forecasting complexity. Retaining too few modes would excessively smooth the wake and remove intermediate-scale coherent structures that are relevant for the subsequent generative reconstruction. Conversely, retaining too many modes would increase the dimensionality of the coefficient space and make the temporal forecasting problem less stable. In the present dataset, retaining 128 modes was found to preserve the dominant wake organization together with a significant portion of the intermediate-energy modal content, while still keeping the learning problem substantially smaller than the original full-field representation. The resulting reduced micro-state becomes

$$\mathbf{a}_t \in \mathbb{R}^{128}, \tag{16}$$

where $\mathbf{a}_t$ denotes the retained POD coefficient vector at time $t$. G-LED is then applied to this coefficient-based representation rather than to the full flow snapshot.

A further encoder inside G-LED maps the retained coefficient vector to a lower-dimensional macro representation, given as

$$\mathbf{z}_t \in \mathbb{R}^{32}, \tag{17}$$

The two reduction steps serve different purposes. The 128 POD coefficients define the retained reduced-order space in which the final reconstruction is performed. This space is selected to preserve the dominant wake organization together with part of the intermediate- and lower-energy structures. By contrast, the 32-dimensional encoded representation is the macro-state evolved by the Transformer in G-LED. This second reduction provides a smoother and more tractable representation for temporal forecasting.

Retaining only 32 POD modes would not be equivalent to the present formulation. In that case, the information contained in modes 33-128 would be permanently discarded before training and could not be reconstructed later. In the proposed formulation, these modes remain part of the reconstruction target within the retained POD space, while their temporal contribution is inferred by the diffusion decoder conditioned on the predicted macro-state. Thus, the internal encoder is not intended to replace POD, but to define the coarse macro-dynamics used by G-LED to forecast the retained POD representation.

The Transformer predicts the temporal evolution of the macro-state, and the diffusion decoder reconstructs the corresponding 128-dimensional POD coefficient vector. The final flow field is then obtained by applying the inverse POD transformation using the retained POD modes. In the POD-guided configurations, both quantitative and qualitative comparisons are performed against the POD-reconstructed reference fields associated with the retained modes, rather than against the original full-field snapshots. Accordingly, the global POD-G-LED configuration should be evaluated as a reduced-order forecasting model within the retained POD subspace, rather than as a full-field reconstruction model.

**2.4. Localized POD-G-LED method**

The second POD-based configuration accounts for the fact that the wake is not spatially uniform in complexity. Close to the cylinder, coherent structures are stronger and more directly represented by energetic POD modes. Farther downstream, the wake becomes more irregular, and a larger fraction of the dynamics is encoded in low-energy content. Therefore, a single global POD basis may not be optimal for representing both regions with the same reduced-order resolution.

To address this issue, the spatial domain is partitioned into a near-cylinder region and a far-wake region, as illustrated in Fig. 4. Following the analysis of Ref. [22], the interface is placed at $x$=17 mm. This location approximately separates the region dominated by stronger coherent structures from the downstream region where the wake becomes more irregular and increasingly dependent on lower-energy modal content. The selected partition is therefore used as a practical and physically motivated division point for testing whether spatially localized reduced-order representations can improve POD-G-LED predictions.

POD and G-LED are then applied independently in each subdomain. In the near-cylinder region, 64 POD modes are retained to preserve a richer representation of the organized vortical structures close to the body. In the far-wake region, 16 POD modes are retained, prioritizing prediction stability and large-scale coherence over detailed small-scale reconstruction. These modal numbers are selected as a compromise between local reconstruction quality and forecasting stability. Retaining more modes near the cylinder helps preserve coherent structures that remain relatively predictable, whereas using fewer modes downstream reduces the influence of highly irregular low-energy coefficients that are more difficult to forecast reliably. Although the present work uses a fixed interface location and prescribed modal numbers, this setting provides a practical first assessment of spatially localized POD-G-LED forecasting. Future work may further optimize these choices using criteria based on local modal energy, reconstruction error, or prediction reliability.

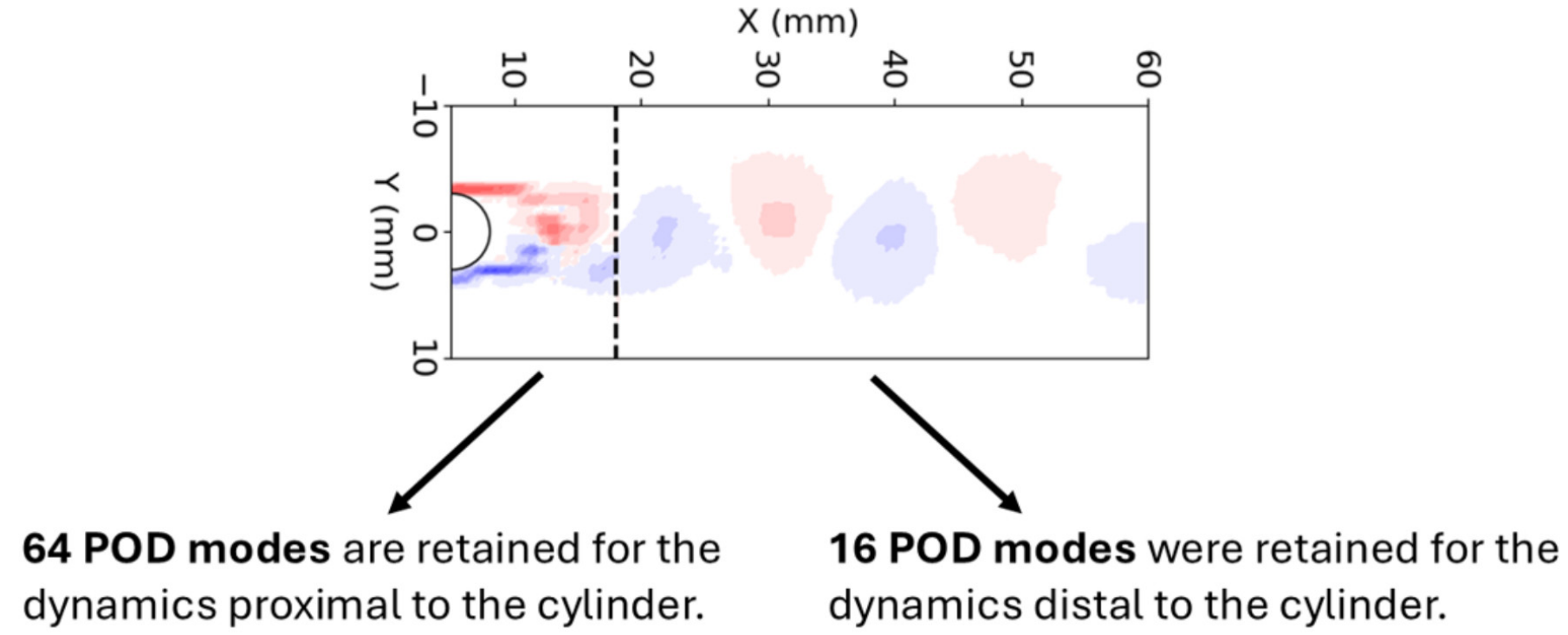

**Fig. 4.** Illustration of the spatial domain partitioned into two disjoint subdomains, within which POD and G-LED methodologies are independently applied.

This localized decomposition introduces two important features. First, the effective reduced basis is allowed to vary with the local complexity of the flow. Second, POD filtering is treated not as a uniform operation over the whole domain, but as a regional design choice that can be adapted to the predictability of local wake dynamics. In this way, the localized POD-G-LED method provides a more flexible reduced-order representation than the global POD-G-LED configuration.

**2.5. Evaluation metrics and physical diagnostics**

Although the models are trained on velocity or POD coefficients, the main evaluation is performed in the vorticity field because vorticity is more sensitive to the presence or absence of small-scale structures. The coefficient of determination is used as the primary global metric

$$R^2 = 1 - \frac{\sum_i (\omega_i - \hat{\omega}_i)^2}{\sum_i (\omega_i - \bar{\omega}_i)^2}, \tag{18}$$

where $\omega_i$ is the reference vorticity, $\hat{\omega}_i$ is the predicted vorticity, and $\bar{\omega}_i$ is the spatial mean of the reference vorticity.

Two additional profile-based diagnostics are employed. The first is the streamwise profile of the root mean square (RMS) vorticity, which measures fluctuation intensity and is particularly informative about intermediate and fine-scale activity. The second is the mean vorticity profile along the streamwise direction, which measures whether the model preserves the average organization of the wake.

Finally, the turbulent kinetic energy (TKE) is computed as a physical consistency check. Since the current dataset consists of two-dimensional velocity fields with two measured velocity components, TKE is evaluated as

$$TKE(x,y) = \frac{1}{2}\left(\overline{u'(x,y,t)^2} + \overline{v'(x,y,t)^2}\right), \tag{19}$$

where $u' = u - \overline{u}$ and $v' = v - \overline{v}$ are the streamwise and transverse velocity fluctuations, respectively. Agreement in TKE does not guarantee exact pointwise reconstruction, but it indicates whether the predicted field preserves the energetic organization of the turbulence in a meaningful way.

## 3. Results

### 3.1. Full-field G-LED prediction

In this section, G-LED is trained directly on velocity snapshots, using the macro encoder only as an internal part of the generative architecture rather than as an external POD filter. The $R^2$ evolution shown in Fig. 5 indicates that the model maintains good predictive skill across the 100-step rollout. Although some degradation is visible as the horizon increases, the model remains capable of tracking the overall wake organization.

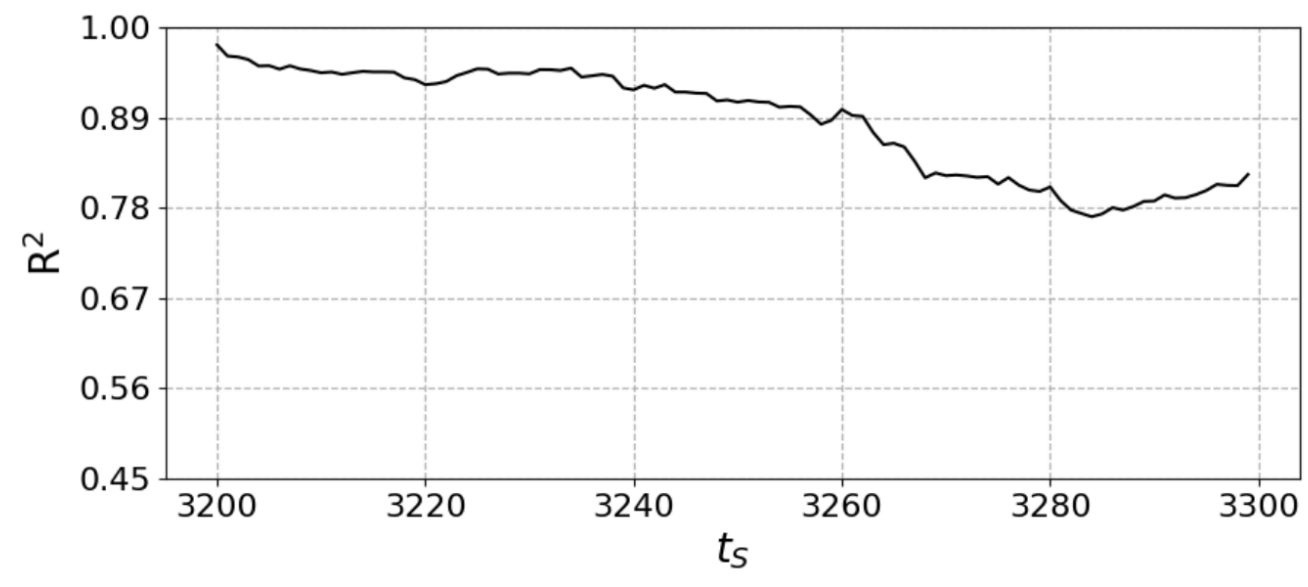

**Fig. 5.** Value of the $R^2$ metric for each prediction generated by G-LED.

Fig. 6 compares the predicted and reference RMS vorticity and mean vorticity profiles. The RMS profile is followed closely, indicating that the model reproduces a substantial part of the fluctuation energy along the wake. The mean vorticity profile is also well captured, which shows that the dominant wake organization is preserved alongside the fluctuations.

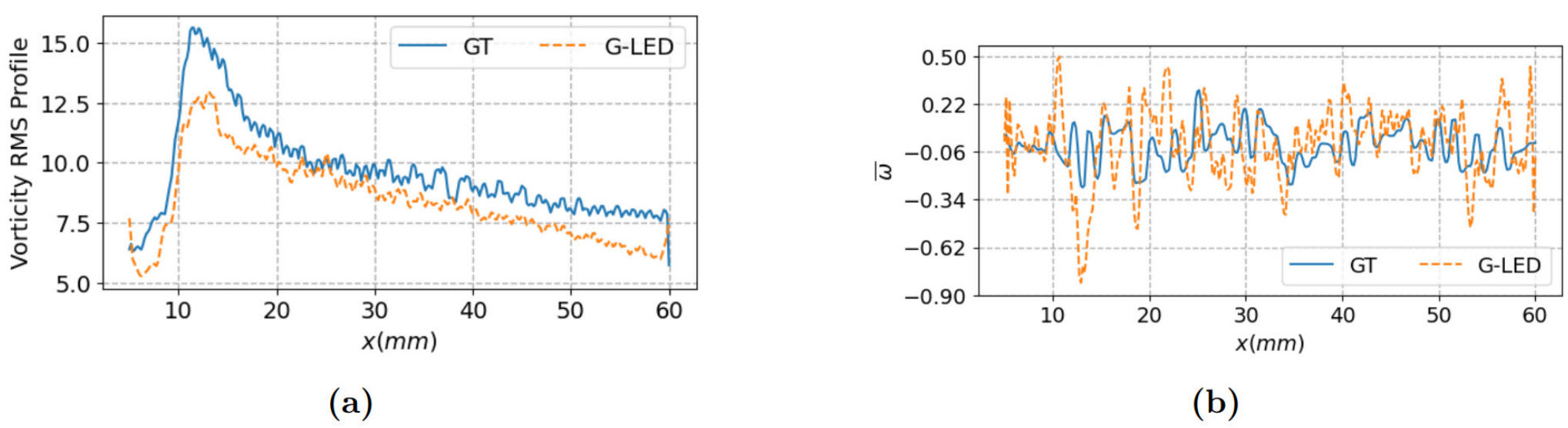

**Fig. 6.** Comparison between the reference (GT) and predicted root-mean-square (RMS) vorticity profile along the streamwise direction (a) and the mean vorticity profile along the streamwise direction (b) for the full-field G-LED configuration.

The most striking evidence appears in Fig. 7, where predicted and reference vorticity snapshots are compared at three representative time instants. The prediction preserves a wide range of localized vortical structures and produces fields that visually resemble the measured turbulent wake. This does not imply exact recovery of every instantaneous fine-scale feature. Such pointwise recovery is not expected for turbulent dynamics because of their chaotic and strongly sensitive nature. The relevant objective is instead to reproduce the dominant coherent structures and the statistical organization of the wake. Nevertheless, the comparison shows that the model captures the qualitative richness of the turbulent field.

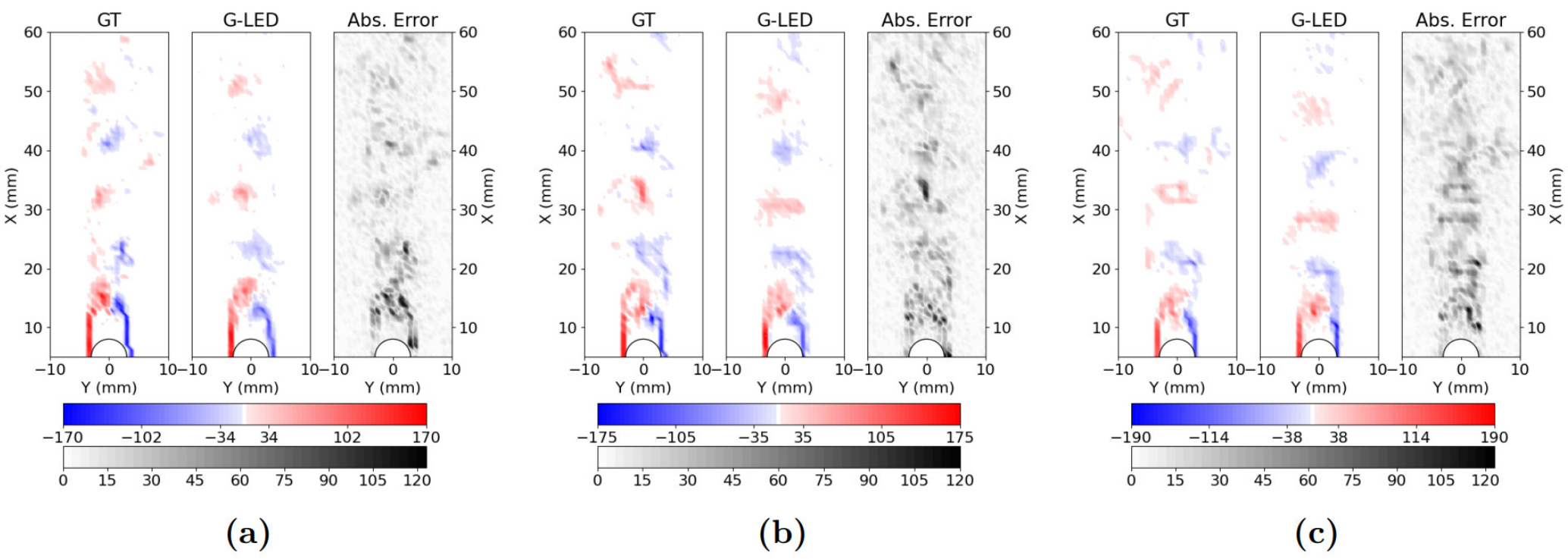

**Fig. 7.** Comparison between the reference (GT) and predicted vorticity snapshots for snapshots (a) 3210,

(b) 3250, and (c) 3300 from the second steady regime for the full-field G-LED configuration.

Fig. 8 further supports this conclusion through the TKE comparison. The predicted energetic region downstream of the cylinder matches the reference distribution reasonably well, and the absolute error remains localized rather than globally spread over the domain. This indicates that the model is not only generating visually plausible structures, but also preserving physically meaningful energetic organization.

These results establish the full-field G-LED configuration as the high-fidelity reference case in the study. At the same time, these results show that this fidelity is obtained at a significant computational cost, which motivates the POD-guided variants.

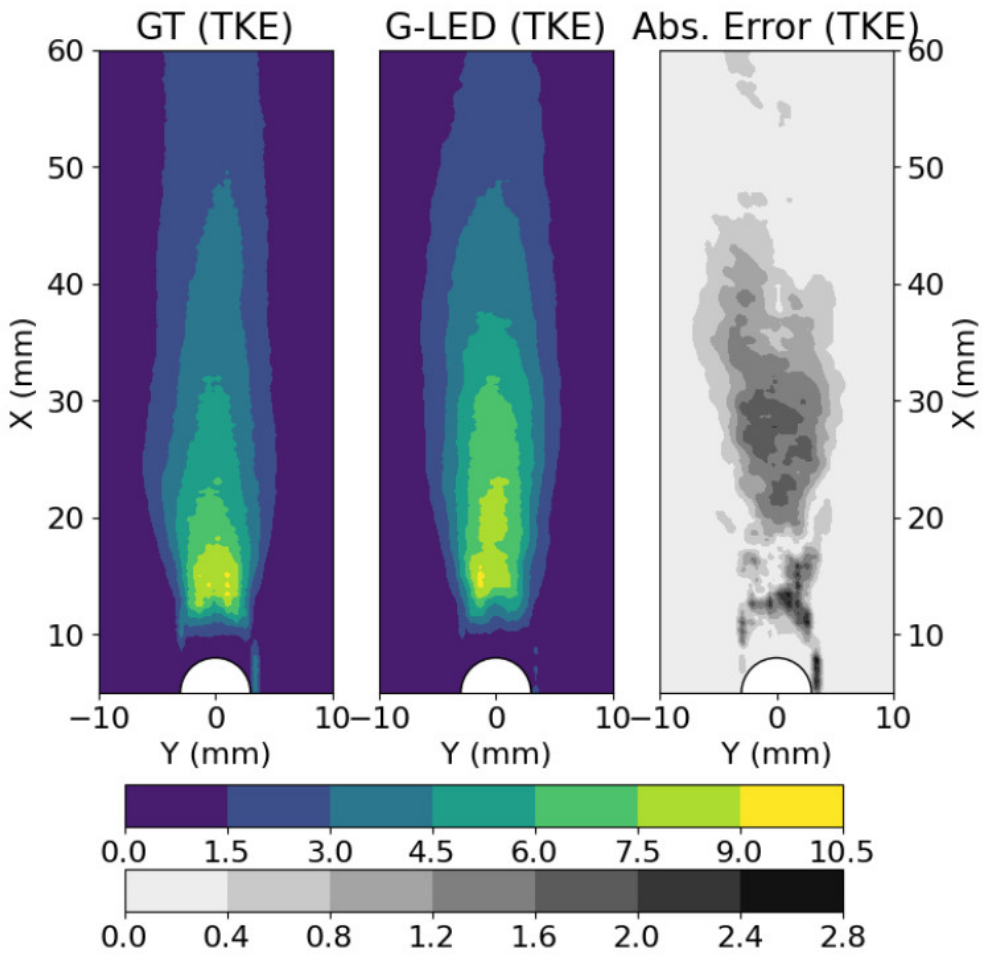

**Fig. 8.** Comparison of the TKE computed from the reference field (left) and from the full-field G-LED prediction (middle). The right column shows the absolute error between prediction and reference.

### 3.2. Prediction of global POD-G-LED

When G-LED is trained on the retained 128-dimensional POD coefficient vector rather than on the full flow fields, the forecasting task is fundamentally reformulated. Instead of directly evolving high-dimensional velocity snapshots, the model operates in a compact reduced-order space. It can preserve the dominant and intermediate flow organization while filtering out the most irregular fine-scale content. This representation offers two advantages. First, the dimensionality of the learning problem is drastically reduced, which lowers the burden on both the Transformer and the diffusion model. Second, the retained

POD coefficients provide a physically interpretable latent description of the wake, making the forecasting process more structured than direct full-field prediction.

Under this formulation, the purpose of global POD-G-LED is not to reproduce every small-scale feature contained in the original wake, but to efficiently predict the filtered wake dynamics within the retained POD subspace. From this perspective, the results remain meaningful. As shown in Fig. 9, the model maintains reasonable $R^2$ values over the prediction horizon, indicating that the dominant temporal evolution of the reduced-order representation is still captured. Fig. 10 shows the RMS and mean vorticity profiles, in which the reference field corresponds to the POD-reconstructed field obtained from the retained modes. It indicates that acceptable agreement is produced over most of the domain, which suggests that the model preserves an important part of the fluctuation level represented in the retained POD space.

The main discrepancy appears in the mean vorticity and in the downstream wake, where the prediction departs more visibly from the reference. However, this behavior should be interpreted together with the role of POD truncation. In the present configuration, the reference field itself is reconstructed from the retained POD modes; therefore, the task is already focused on the prediction of a filtered and physically meaningful surrogate of the flow. In this sense, the global POD-G-LED results demonstrate that generative forecasting remains feasible even after modal compression, and that a substantial part of the wake organization can still be recovered at a much lower computational cost than in the full-field configuration.

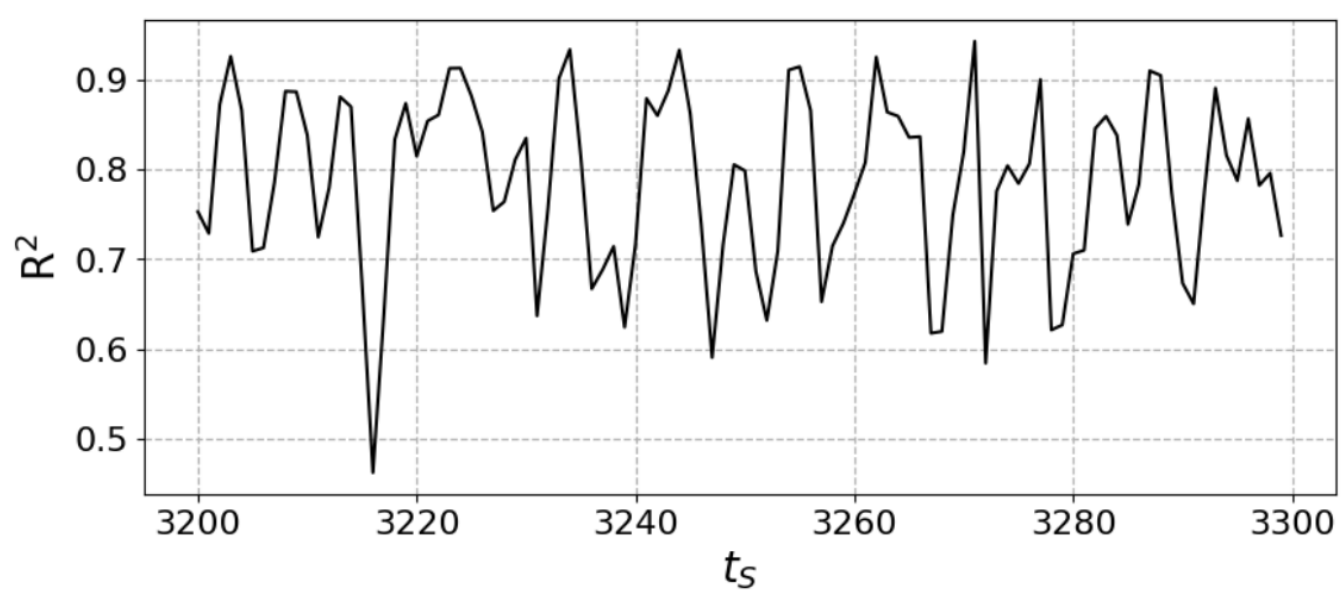

**Fig. 9.** $R^2$ metric values for each prediction generated by global POD-G-LED.

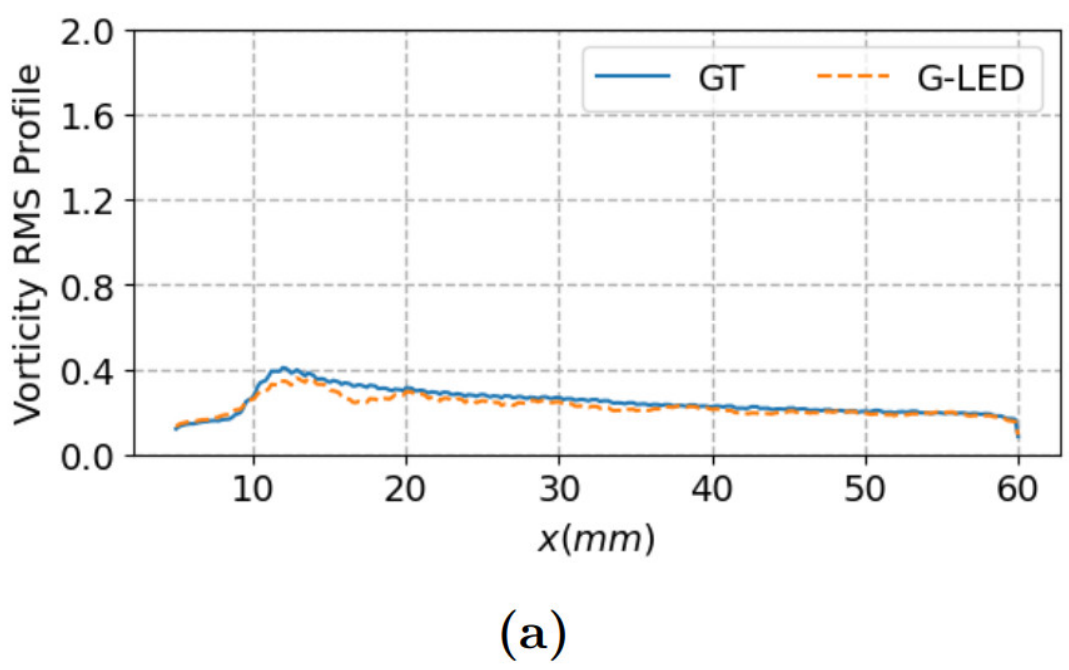

(a)

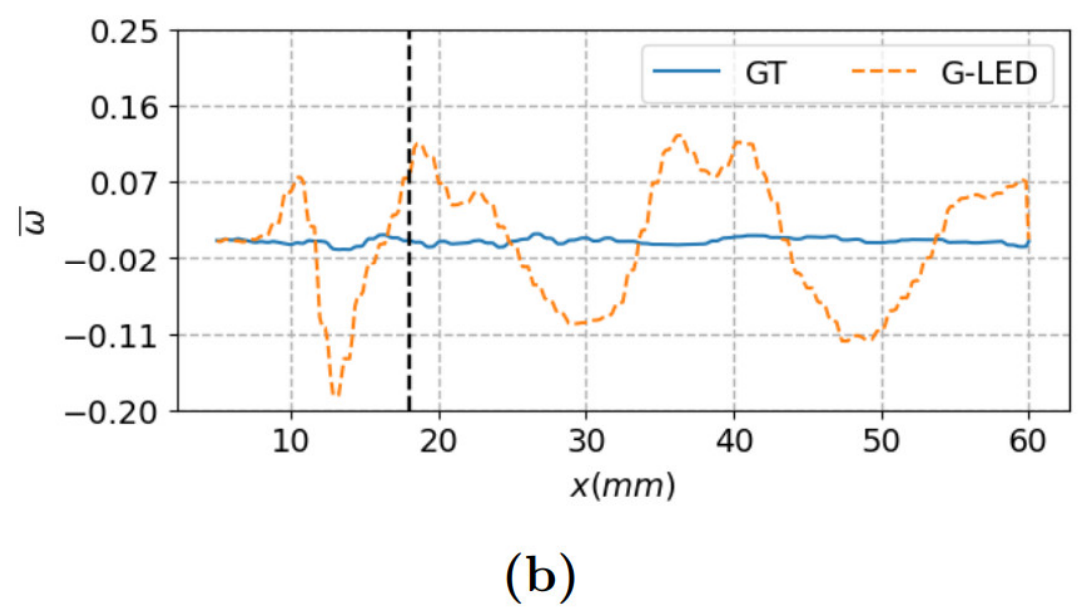

(b)

**Fig. 10.** Comparison between the POD-reconstructed reference (GT) and the global POD-G-LED predictions for the root-mean-square (RMS) vorticity profile along the streamwise direction (a) and the mean vorticity profile along the streamwise direction (b).

This point is also reflected in the instantaneous fields shown in Fig. 11. Although the reconstructed wake is smoother than in the direct full-field G-LED case, the main vortical organization is preserved, especially in the near-cylinder region. Similarly, the TKE maps in Fig. 12 show that the global POD-G-LED model still captures the main energetic distribution of the wake, although the TKE intensity is underestimated in regions dominated by finer-scale structures. Therefore, the value of global POD-G-LED lies in its ability to couple a compact POD representation with diffusion-based generative forecasting, producing stable, interpretable, and computationally efficient predictions of the dominant coherent wake structures and their temporal evolution.

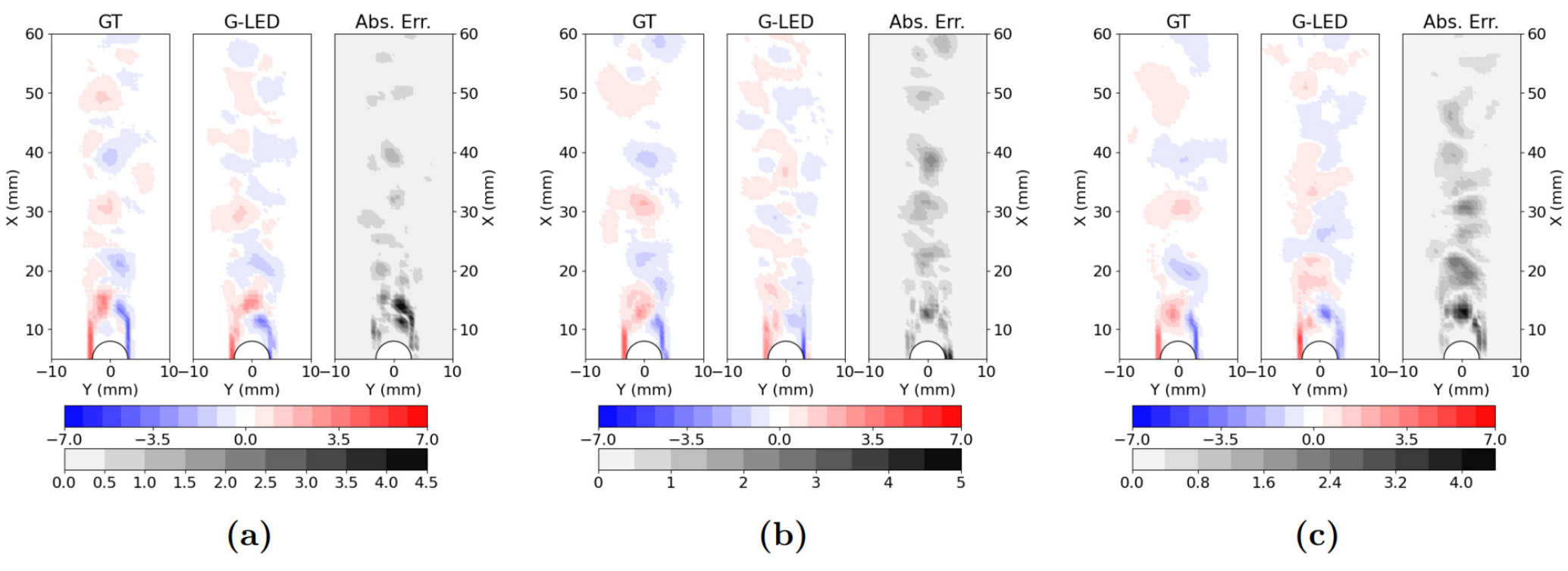

(a) (b) (c)

**Fig. 11.** Comparison between the POD-reconstructed reference (GT) and the predicted vorticity snapshots for snapshots (a) 3210, (b) 3250, and (c) 3300 from the second steady regime for the global POD-G-LED configuration.

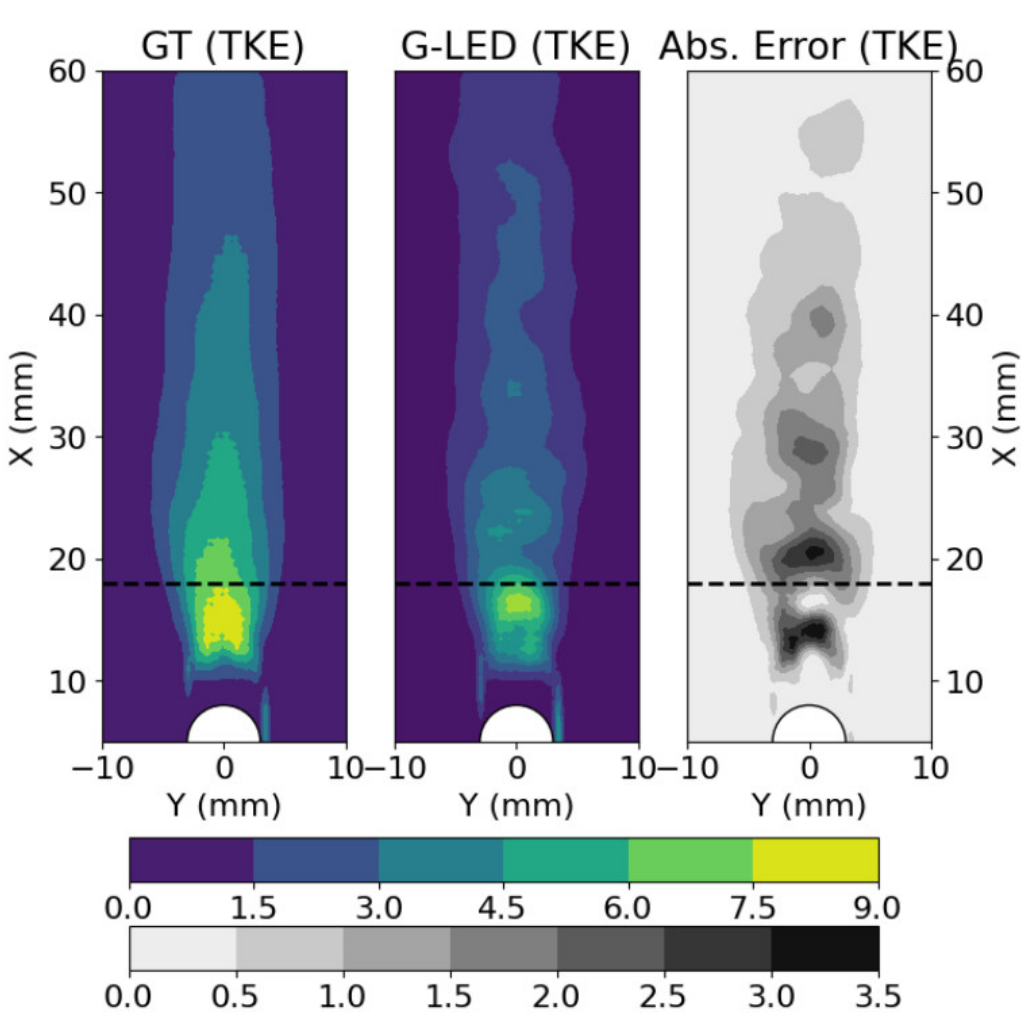

**Fig. 12.** Comparison of the TKE computed from the POD-reconstructed reference field (left) and from the global POD-G-LED prediction (middle). The right column shows the absolute error between prediction and reference.

The significance of this result is therefore twofold. On one hand, it confirms that POD filtering substantially reduces the learning burden: the diffusion-model training time decreases from about 17 h to 8 h, the Transformer training time from about 7 h to 2 h, and the inference time for 100 future predictions from about 3 min to 50 s. On the other hand, it shows that diffusion-based generative forecasting can still add value in the reduced-order setting by improving the prediction of retained POD dynamics that are too irregular for standard deterministic POD-based predictors. Global POD-G-LED should thus be understood as an efficiency-oriented surrogate strategy, particularly relevant when the target is the dominant and intermediate wake organization rather than the complete recovery of all fine-scale turbulent content.

### 3.3. Prediction of localized POD-G-LED

The results of the global POD-G-LED configuration suggest that a single global reduced basis may not be sufficient for a wake whose spatial complexity is strongly non-uniform. In particular, the near-cylinder region is dominated by more coherent structures that can still benefit from a richer modal representation, whereas the farther wake contains increasingly irregular dynamics that are less efficiently represented in a single global POD basis. This observation motivates the localized POD-G-

LED strategy.

In the localized approach, the domain is divided into two subregions and modeled independently with different modal resolutions. This modification is important not only as a numerical adjustment, but also as a conceptual extension of POD-guided generative forecasting. It shows that POD filtering does not need to be applied uniformly over the whole domain; rather, the reduced-order resolution can be adapted to the local predictability and structural complexity of the wake. In this sense, the localized POD-G-LED framework better reflects the spatially heterogeneous nature of turbulent wakes.

The effectiveness of this strategy is evident in the results. As shown in Fig. 13, the $R^2$ values remain high throughout the rollout, indicating robust temporal prediction of the localized reduced-order representation. Fig. 14 shows that both the RMS and mean vorticity profiles exhibit improved agreement with the reference compared with the global POD-G-LED case. The most noticeable discrepancy is concentrated around the 17 mm location, where the two independently reconstructed subdomains are merged. This interface artifact is expected, since two local reduced-order models are coupled there. Importantly, away from this narrow transition region, the agreement is clearly improved.

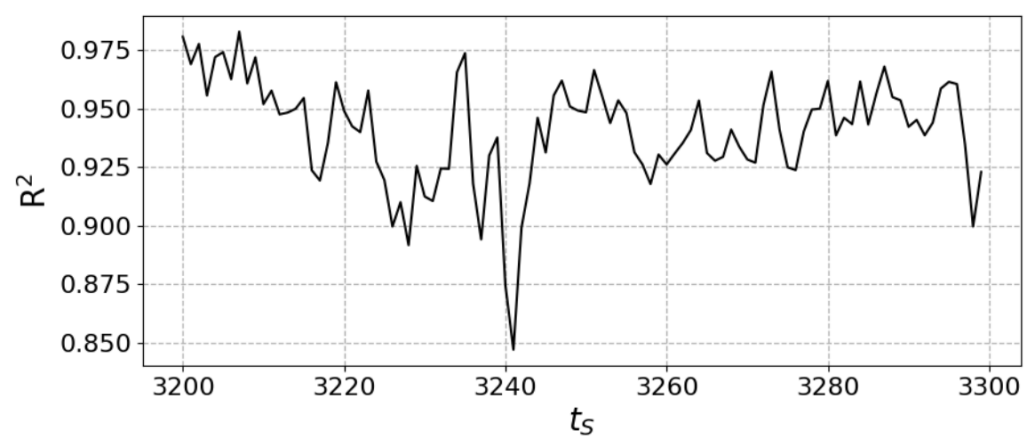

**Fig. 13.** $R^2$ metric values for each prediction generated by the localized POD-G-LED configuration.

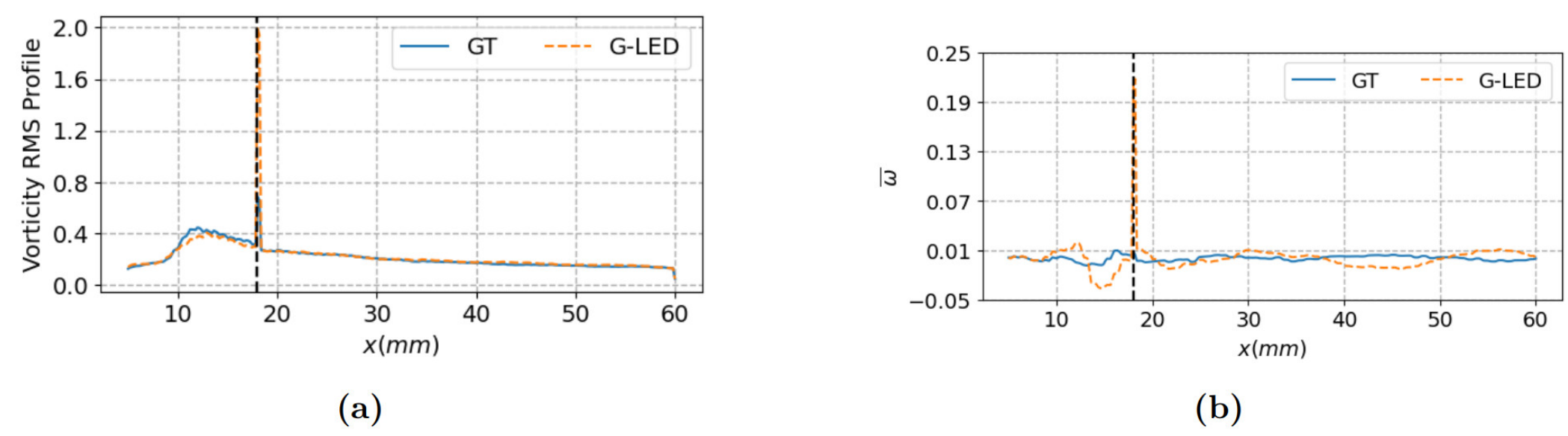

**Fig. 14.** Comparison between the POD-reconstructed reference (GT) and the localized POD-G-LED predictions for the root-mean-square (RMS) vorticity profile along the streamwise direction (a) and the mean vorticity profile along the streamwise direction (b).

The instantaneous vorticity fields in Fig. 15 further highlight the value of the localized approach. Compared with the global POD-G-LED case, the near-cylinder region retains richer vortical detail, indicating that the larger local modal basis preserves more of the organized structures that dominate this part of the wake. At the same time, the far-wake region remains stable and coherent despite the use of a more compact representation. Away from the interface, the localized reconstruction performs substantially better than the global POD-G-LED case.

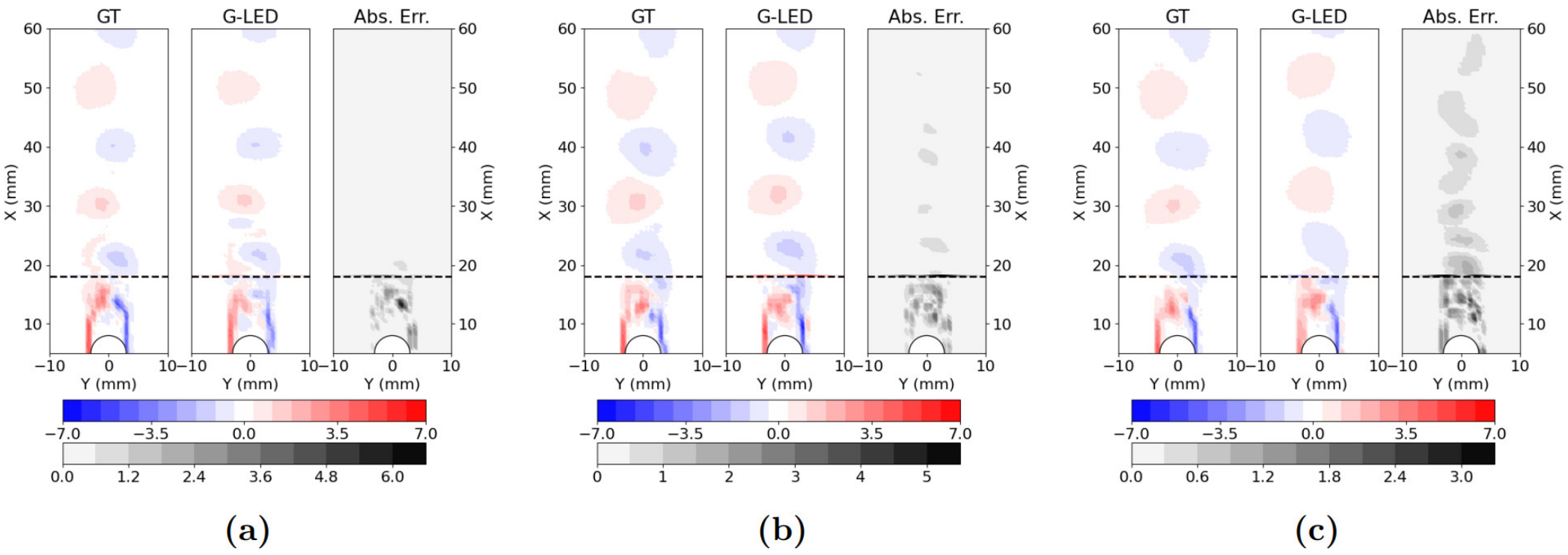

**Fig. 15.** Comparison between the POD-reconstructed reference (GT) and the predicted vorticity snapshots for snapshots (a) 3210, (b) 3250, and (c) 3300 from the second steady regime for the localized POD-G-LED configuration.

The same conclusion is supported by the TKE distribution in Fig. 16. The localized POD-G-LED prediction is more physically consistent than that of the global POD-G-LED configuration, indicating that regional modal adaptation improves not only visual reconstruction but also the preservation of the energetic organization of the flow. This is a key result, because it shows that the limitations of POD-guided generative forecasting do not arise solely from POD truncation itself, but also from how that truncation is distributed in space.

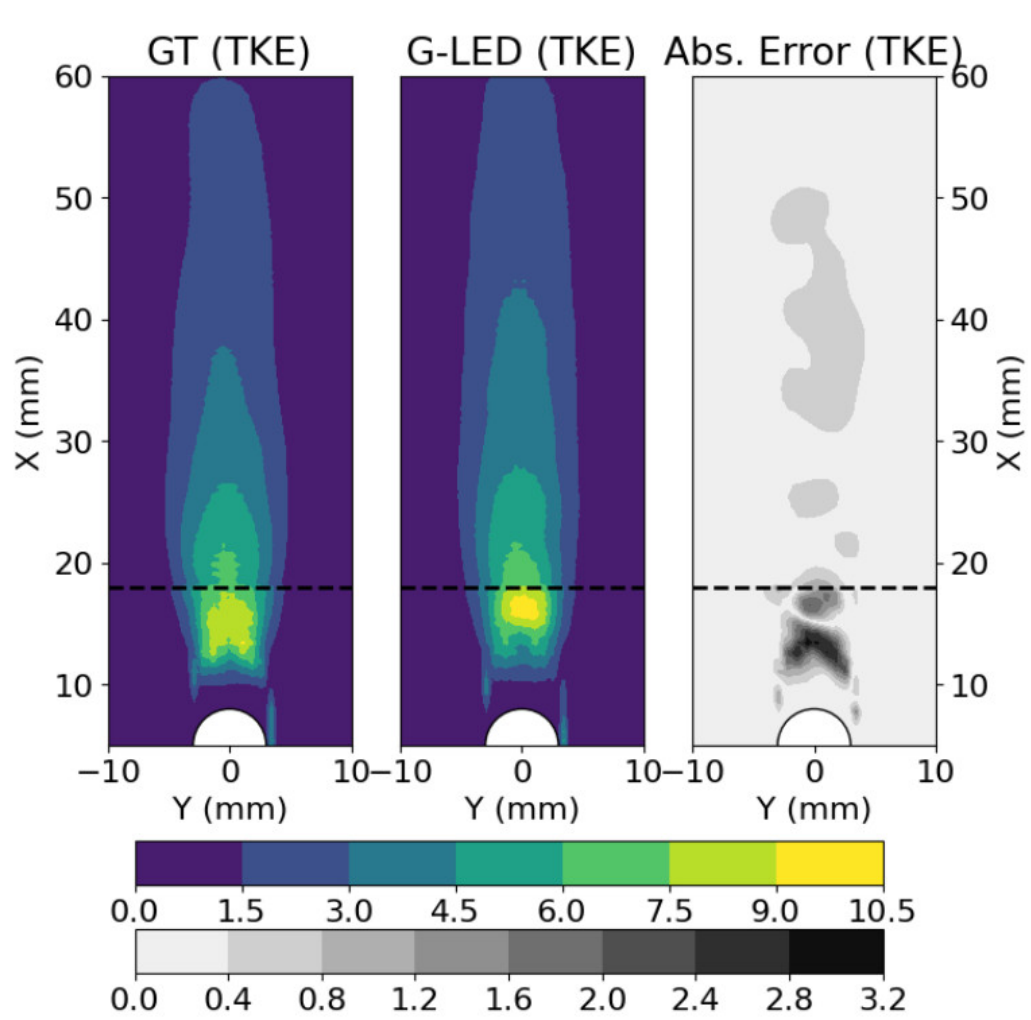

**Fig. 16.** Comparison of the TKE computed from the POD-reconstructed reference field (left) and from the localized POD-G-LED prediction (middle). The right column shows the absolute error between prediction and reference.

The localized strategy therefore has clear research value. It demonstrates that the combination of POD and G-LED should not be viewed as a single fixed pipeline, but as a flexible reduced-order generative framework whose representation can be tailored to the flow structure. Although it does not recover the same level of detail as the direct full-field G-LED model, it provides a substantially better fidelity–efficiency compromise than the global POD-G-LED configuration. More importantly, it establishes that POD-guided generative forecasting can be improved in a physically meaningful way by introducing spatially adaptive reduced-order representations, which is highly relevant for complex turbulent flows with non-uniform spatial organization.

### 3.4. Computational efficiency analysis

A comparison of training and inference costs was carried out to evaluate the computational advantage of POD-guided generative forecasting. For the full-field G-LED configuration, training was performed on two RTX A4500 GPUs, one assigned to the Transformer and the other to the diffusion model. Under this setting, the diffusion model required approximately 17 h for training, while the Transformer required about 7 h. The prediction of 100 future flow fields then required about 3 min on a single RTX A4500 GPU.

In the POD-guided configurations, the learning task was transferred from high-dimensional flow snapshots to low-dimensional POD coefficient vectors. Under the same hardware conditions, the diffusion-model training time decreased to approximately 8 h, whereas the Transformer training time was reduced to about 2 h. Inference for 100 future predictions was also reduced to approximately 50 s. Therefore, relative to the full-field configuration, the POD-guided strategy provides a marked reduction in both training and inference cost.

These results highlight the practical value of introducing POD prior to generative forecasting. Although some loss of fine-scale fidelity is observed, the resulting efficiency gain makes the POD-guided framework particularly attractive for applications in which rapid prediction, repeated model updates, or reduced deployment cost are of primary importance.

## 4. Discussion

The present results show that the combination of POD and G-LED should not be interpreted merely as a simplified version of the full-field generative model. Instead, it defines a reduced-order generative forecasting framework in which modal compression, low-dimensional temporal prediction, and stochastic reconstruction are explicitly linked. Within this framework, POD provides a compact and physically meaningful representation of the flow dynamics, whereas G-LED extends the predictive capability of this reduced-order space by coupling low-dimensional forecasting with diffusion-based reconstruction. From this perspective, the proposed methodology occupies an intermediate position between conventional POD-based forecasting and direct full-field generative prediction.

The full-field G-LED configuration serves as the high-fidelity reference. Since it operates directly on velocity snapshots, it is able to reconstruct richer fluctuation content and more detailed vortical structures. This is reflected by the better agreement in the vorticity fields, RMS vorticity profiles, mean vorticity profiles, and TKE distributions. However, this higher fidelity is obtained at the cost of substantially higher training and inference times. Therefore, full-field G-LED is most appropriate when the main objective is to preserve as much flow-field detail as possible and the associated computational cost is acceptable.

By contrast, the POD-guided configurations pursue a different objective. They do not aim to recover all unresolved small-scale content present in the original flow field. Their purpose is to predict the retained reduced-order dynamics as efficiently as possible while preserving the dominant coherent structures that govern the temporal evolution of the wake. Introducing POD before G-LED greatly reduces the training cost of both the diffusion model and the Transformer, and also substantially decreases the inference time. This confirms that POD is not merely a preprocessing step, but a central component that changes the computational scale of the generative forecasting problem.

The results clarify the limitations of a global reduced-order representation. In the turbulent cylinder wake considered here, the flow is not spatially uniform in complexity. The near-cylinder region is dominated by more coherent and energetic structures, whereas the farther wake contains increasingly irregular dynamics that are encoded to a larger extent in lower-energy content. A single global POD basis cannot adapt equally well to both regions. This explains why the global POD-G-LED configuration, although efficient, shows a reduction in fidelity, especially farther downstream. Importantly, this does not imply that POD-guided generative forecasting is ineffective; rather, it indicates that the way the reduced-order representation is constructed becomes a critical design choice.

This observation motivates the localized POD-G-LED strategy, which represents one of the most meaningful outcomes of the study. Its significance is not limited to the improved quantitative results relative to the global POD-G-LED case. More fundamentally, it shows that reduced-order generative forecasting does not need to rely on a single global representation of the flow. By allowing the number of retained modes to vary across the domain, the localized strategy adapts the reduced-order resolution to the local structure and predictability of the wake. In the present work, this idea is implemented through a simple partition into two subdomains with different modal resolutions. Even with this relatively simple design, the localized configuration already provides a markedly better fidelity-efficiency compromise than the global POD-G-LED case. This suggests that spatial adaptivity is likely to be an important ingredient in future reduced-order generative models for complex turbulent flows.

The implications of this result extend beyond the specific cylinder wake examined here. More generally, it suggests that the coupling between reduced-order modeling and generative reconstruction should not necessarily be uniform in space. Instead, it can be designed to reflect local differences in coherence, scale content, and predictability. In this sense, the localized POD-G-LED approach points toward a broader class of spatially adaptive reduced-order generative models, in which modal truncation and generative reconstruction are adjusted to the heterogeneous nature of the flow. Such a perspective may be particularly relevant for turbulent configurations where different regions of the domain exhibit markedly different dynamical behavior.

Overall, the present study indicates that POD-guided G-LED should be regarded as a flexible and practically meaningful reduced-order generative framework. Its main value does not lie in fully reconstructing all unresolved small-scale turbulence, but in efficiently forecasting a filtered and physically interpretable representation of the wake while preserving a significant portion of its dominant and intermediate-scale organization. Under this interpretation, the method is not a weakened version of full-field G-LED, but a complementary strategy with a different target: improved efficiency, interpretability, and adaptability within a reduced-order generative setting. This makes the framework particularly relevant for surrogate-assisted studies in which dominant and intermediate-scale dynamics are of greater practical importance than exact recovery of all fine-scale turbulent structures.

## 5. Conclusions

This work proposed a hybrid reduced-order generative forecasting framework that combines proper orthogonal decomposition (POD) with the Generative Learning of Effective Dynamics (G-LED) model for the prediction of turbulent cylinder wakes. The central idea is to perform temporal forecasting in a low-dimensional physics-based modal space and subsequently reconstruct physically meaningful flow fields through diffusion-based generative modeling. Three configurations were assessed using the same experimental dataset: full-field G-LED, global POD-G-LED, and localized POD-G-LED.

Three main conclusions can be drawn. First, the full-field G-LED configuration provides the highest prediction fidelity. It better preserves fluctuation-rich vorticity structures, reproduces RMS-vorticity distributions more accurately, and yields turbulent kinetic energy fields that are closer to the experimental reference. This level of fidelity, however, comes at a significant computational cost, requiring approximately 17 h for diffusion-model training, 7 h for Transformer training, and about 3 min to predict 100 future snapshots.

Second, introducing POD prior to generative forecasting substantially reduces the computational burden by transferring the learning task from high-dimensional flow fields to low-dimensional modal coefficients. In the global POD-G-LED configuration, diffusion-model training is reduced to approximately 8 h, Transformer training to 2 h, and inference time for 100 future predictions to about 50 s. Despite this reduction, the reconstructed flow fields retain the dominant wake organization and the most energetic coherent structures with good accuracy. These results demonstrate that coupling physics-based modal representations with diffusion-based reconstruction provides an effective compromise between computational efficiency and predictive fidelity.

Third, the localized POD-G-LED strategy provides a better balance between accuracy and efficiency among the reduced-order approaches. By assigning different modal resolutions to different wake regions, the method better captures the spatially non-uniform dynamics of the flow, improving vorticity statistics and energetic distributions relative to the global POD-G-LED formulation. This finding highlights the importance of adapting the reduced-order representation to local flow complexity and suggests that spatial adaptivity is a key ingredient for future generative reduced-order models.

The present study was restricted to a single experimental cylinder-wake dataset, and the localized formulation relied on a prescribed interface location and fixed modal resolutions. Future work will focus on adaptive domain partitioning, data-driven modal-selection strategies, and uncertainty quantification exploiting the probabilistic nature of diffusion models. More broadly, the present results indicate that diffusion-enhanced reduced-order modeling provides a promising pathway toward bridging the gap between computationally efficient flow representations and high-fidelity reconstruction of turbulent-flow dynamics.

**Acknowledgment**

The authors acknowledge the funding from the European Union's Horizon Europe research and innovation programme under the Marie Skłodowska-Curie grant agreements ENCODING No. 101072779, and MODELAIR No. 101072559. Views and opinions expressed are however those of the author(s) only and do not necessarily reflect those of the European Union or the European Research Executive Agency. Neither the European Union nor the granting authority can be held responsible for them. The authors also acknowledge the grant PID2023-147790OB-I00 funded by MCIU/AEI/10.13039/501100011033/FEDER, UE. The authors gratefully acknowledge the Universidad Politécnica de Madrid (www.upm.es) and CeSViMa for providing computing resources on the Magerit Supercomputer.

**Competing interests**

The authors have no relevant financial or non-financial interests to disclose.

**Data availability statement**

The data that support the findings of this study are available from the corresponding author upon reasonable request.

**References**

[1] Brunton SL, Noack BR, Koumoutsakos P. Machine learning for fluid mechanics. Annu Rev Fluid Mech 2020;52:477–508.

[2] Vinuesa R, Brunton SL. Enhancing computational fluid dynamics with machine learning. Nat Comput Sci 2022;2:358–66. https://doi.org/10.1038/s43588-022-00264-7.

[3] Ihme M, Chung WT, Mishra AA. Combustion machine learning: Principles, progress and prospects.

Prog Energy Combust Sci 2022;91:101010. https://doi.org/10.1016/j.pecs.2022.101010.

[4] Price I, Sanchez-Gonzalez A, Alet F, Andersson TR, El-Kadi A, Masters D, et al. Probabilistic weather forecasting with machine learning. Nature 2025;637:84–90. https://doi.org/10.1038/s41586-024-08252-9.

[5] An J, Wang H, Liu B, Luo KH, Qin F, He GQ. A deep learning framework for hydrogen-fueled turbulent combustion simulation. Int J Hydrog Energy 2020;45:17992–8000. https://doi.org/10.1016/j.ijhydene.2020.04.286.

[6] Zou X, Parente A, Le Clainche S. Divergence detection and flow structure analysis in POD-DL predictions of a hydrogen-methane flame. Eur J Mech - BFluids 2026;119:204515. https://doi.org/10.1016/j.euromechflu.2026.204515.

[7] Zou X, Abadia-Heredia R, Saavedra L, Parente A, Xue R, Clainche SL. Generative artificial intelligence and hybrid models to accelerate LES in reactive flows: Application to hydrogen/methane combustion. ArXiv Prepr ArXiv250708426 2025.

[8] Yang Y, Jin M, Wen H, Zhang C, Liang Y, Ma L, et al. A survey on diffusion models for time series and spatio-temporal data. ACM Comput Surv 2026;58:1–39. https://doi.org/10.1145/3783986.

[9] Lin L, Li Z, Li R, Li X, Gao J. Diffusion models for time-series applications: a survey. Front Inf Technol Electron Eng 2024;25:19–41. https://doi.org/10.1631/FITEE.2300310.

[10] Cachay SR, Zhao B, Joren H, Yu R. DYffusion: A Dynamics-informed Diffusion Model for Spatiotemporal Forecasting 2023;36:45259–87.

[11] Lienen M, Lüdke D, Hansen-Palmus J, Günnemann S. From Zero to Turbulence: Generative Modeling for 3D Flow Simulation 2024. https://doi.org/10.48550/arXiv.2306.01776.

[12] Du P, Parikh MH, Fan X, Liu X-Y, Wang J-X. Conditional neural field latent diffusion model for generating spatiotemporal turbulence. Nat Commun 2024;15:10416. https://doi.org/10.1038/s41467-024-54712-1.

[13] Gao H, Kaltenbach S, Koumoutsakos P. Generative learning for forecasting the dynamics of high-dimensional complex systems. Nat Commun 2024;15:8904. https://doi.org/10.1038/s41467-024-53165-w.

[14] Vega JM, Le Clainche S. Higher order dynamic mode decomposition and its applications.

Academic Press; 2020.

[15] Brunton SL, Kutz JN. Data-Driven Science and Engineering : Machine Learning, Dynamical Systems, and Control 2022.

[16] Berkooz G, Holmes P, Lumley JL. The Proper Orthogonal Decomposition in the Analysis of Turbulent Flows. Annu Rev Fluid Mech 1993;25:539–75. https://doi.org/10.1146/annurev.fl.25.010193.002543.

[17] Jacquier P, Abdedou A, Delmas V, Soulaïmani A. Non-intrusive reduced-order modeling using uncertainty-aware Deep Neural Networks and Proper Orthogonal Decomposition: Application to flood modeling. J Comput Phys 2021;424:109854. https://doi.org/10.1016/j.jcp.2020.109854.

[18] Pant P, Doshi R, Bahl P, Barati Farimani A. Deep learning for reduced order modelling and efficient temporal evolution of fluid simulations. Phys Fluids 2021;33:107101. https://doi.org/10.1063/5.0062546.

[19] Bukka SR, Gupta R, Magee AR, Jaiman RK. Assessment of unsteady flow predictions using hybrid deep learning based reduced-order models. Phys Fluids 2021;33:013601. https://doi.org/10.1063/5.0030137.

[20] Abadía-Heredia R, López-Martín M, Carro B, Arribas JI, Pérez JM, Le Clainche S. A predictive hybrid reduced order model based on proper orthogonal decomposition combined with deep learning architectures. Expert Syst Appl 2022;187:115910. https://doi.org/10.1016/j.eswa.2021.115910.

[21] Yang M, Xiao Z. POD-based surrogate modeling of transitional flows using an adaptive sampling in Gaussian process. Int J Heat Fluid Flow 2020;84:108596.

[22] Sengupta A, Abadía-Heredia R, Hetherington A, Perez Perez JM, Le Clainche S. Hybrid machine learning models based on physical patterns to accelerate CFD simulations: a short guide on autoregressive models. Available SSRN 5236454 2025.

[23] Mohan AT, Gaitonde DV. A deep learning based approach to reduced order modeling for turbulent flow control using LSTM neural networks 2018. https://doi.org/10.48550/arXiv.1804.09269.

[24] Nakamura T, Fukami K, Hasegawa K, Nabae Y, Fukagata K. Convolutional neural network and long short-term memory based reduced order surrogate for minimal turbulent channel flow. Phys

Fluids 2021;33:025116. https://doi.org/10.1063/5.0039845.

[25] Mendez MA, Balabane M, Buchlin J-M. Multi-scale proper orthogonal decomposition of complex fluid flows. J Fluid Mech 2019;870:988–1036.

[26] Ronneberger O, Fischer P, Brox T. U-net: Convolutional networks for biomedical image segmentation. Int. Conf. Med. Image Comput. Comput.-Assist. Interv., Springer; 2015, p. 234–41.

[27] Geneva N, Zabaras N. Transformers for modeling physical systems. Neural Netw 2022;146:272–89. https://doi.org/10.1016/j.neunet.2021.11.022.

[28] Sun L, Han X, Gao H, Wang J-X, Liu L. Unifying predictions of deterministic and stochastic physics in mesh-reduced space with sequential flow generative model. Adv Neural Inf Process Syst 2023;36:60636–60.

[29] Ho J, Salimans T, Gritsenko A, Chan W, Norouzi M, Fleet DJ. Video diffusion models. Adv Neural Inf Process Syst 2022;35:8633–46.

[30] Kingma D, Salimans T, Poole B, Ho J. Variational diffusion models. Adv Neural Inf Process Syst 2021;34:21696–707.

[31] De Bortoli V, Thornton J, Heng J, Doucet A. Diffusion schrödinger bridge with applications to score-based generative modeling. Adv Neural Inf Process Syst 2021;34:17695–709.

[32] Gao H, Han X, Fan X, Sun L, Liu L-P, Duan L, et al. Bayesian conditional diffusion models for versatile spatiotemporal turbulence generation. Comput Methods Appl Mech Eng 2024;427:117023.

[33] Ho J, Chan W, Saharia C, Whang J, Gao R, Gritsenko A, et al. Imagen video: High definition video generation with diffusion models. ArXiv Prepr ArXiv221002303 2022.

[34] Song Y, Ermon S. Improved techniques for training score-based generative models. Adv Neural Inf Process Syst 2020;33:12438–48.

[35] Chan S. Tutorial on diffusion models for imaging and vision. Found Trends Comput Graph Vis 2024;16:322–471.

[36] Ho J, Jain A, Abbeel P. Denoising diffusion probabilistic models. Adv Neural Inf Process Syst 2020;33:6840–51.